\documentstyle[twoside,psfig]{article}

\catcode`\@=11
\long\def\@makefntext#1{
\protect\noindent \hbox to 3.2pt {\hskip-.9pt  
$^{{\eightrm\@thefnmark}}$\hfil}#1\hfill}		

\def\thefootnote{\fnsymbol{footnote}}
\def\@makefnmark{\hbox to 0pt{$^{\@thefnmark}$\hss}}	
	
\def\ps@myheadings{\let\@mkboth\@gobbletwo
\def\@oddhead{\hbox{}
\rightmark\hfil\eightrm\thepage}   
\def\@oddfoot{}\def\@evenhead{\eightrm\thepage\hfil
\leftmark\hbox{}}\def\@evenfoot{}
\def\sectionmark##1{}\def\subsectionmark##1{}}

\oddsidemargin=\evensidemargin
\renewcommand{\thefootnote}{\fnsymbol{footnote}}

\newcounter{sectionc}\newcounter{subsectionc}\newcounter{subsubsectionc}
\renewcommand{\section}[1] {\vspace{12pt}\addtocounter{sectionc}{1} 
\setcounter{subsectionc}{0}\setcounter{subsubsectionc}{0}\noindent 
	{\tenbf\thesectionc. #1}\par\vspace{5pt}}
\renewcommand{\subsection}[1] {\vspace{12pt}\addtocounter{subsectionc}{1} 
	\setcounter{subsubsectionc}{0}\noindent 
	{\bf\thesectionc.\thesubsectionc. {\kern1pt \bfit #1}}\par\vspace{5pt}}
\renewcommand{\subsubsection}[1] {\vspace{12pt}\addtocounter{subsubsectionc}{1}
	\noindent{\tenrm\thesectionc.\thesubsectionc.\thesubsubsectionc.
	{\kern1pt \tenit #1}}\par\vspace{5pt}}
\newcommand{\nonumsection}[1] {\vspace{12pt}\noindent{\tenbf #1}
	\par\vspace{5pt}}

\newcounter{appendixc}
\newcounter{subappendixc}[appendixc]
\newcounter{subsubappendixc}[subappendixc]
\renewcommand{\thesubappendixc}{\Alph{appendixc}.\arabic{subappendixc}}
\renewcommand{\thesubsubappendixc}
	{\Alph{appendixc}.\arabic{subappendixc}.\arabic{subsubappendixc}}

\renewcommand{\appendix}[1] {\vspace{12pt}
        \refstepcounter{appendixc}
        \setcounter{figure}{0}
        \setcounter{table}{0}
        \setcounter{lemma}{0}
        \setcounter{theorem}{0}
        \setcounter{corollary}{0}
        \setcounter{definition}{0}
        \setcounter{equation}{0}
        \renewcommand{\thefigure}{\Alph{appendixc}.\arabic{figure}}
        \renewcommand{\thetable}{\Alph{appendixc}.\arabic{table}}
        \renewcommand{\theappendixc}{\Alph{appendixc}}
        \renewcommand{\thelemma}{\Alph{appendixc}.\arabic{lemma}}
        \renewcommand{\thetheorem}{\Alph{appendixc}.\arabic{theorem}}
        \renewcommand{\thedefinition}{\Alph{appendixc}.\arabic{definition}}
        \renewcommand{\thecorollary}{\Alph{appendixc}.\arabic{corollary}}
        \renewcommand{\theequation}{\Alph{appendixc}.\arabic{equation}}
        \noindent{\tenbf Appendix \theappendixc #1}\par\vspace{5pt}}
\newcommand{\subappendix}[1] {\vspace{12pt}
        \refstepcounter{subappendixc}
        \noindent{\bf Appendix \thesubappendixc. {\kern1pt \bfit #1}}
	\par\vspace{5pt}}
\newcommand{\subsubappendix}[1] {\vspace{12pt}
        \refstepcounter{subsubappendixc}
        \noindent{\rm Appendix \thesubsubappendixc. {\kern1pt \tenit #1}}
	\par\vspace{5pt}}

\newcommand{\textlineskip}{\baselineskip=13pt}
\newcommand{\smalllineskip}{\baselineskip=10pt}

\def\eightcirc{
\begin{picture}(0,0)
\put(4.4,1.8){\circle{6.5}}
\end{picture}}
\def\eightcopyright{\eightcirc\kern2.7pt\hbox{\eightrm c}}

\def\abstracts#1#2#3{{
	\centering{\begin{minipage}{4.5in}\footnotesize\baselineskip=10pt
	\parindent=0pt #1\par 
	\parindent=15pt #2\par
	\parindent=15pt #3
	\end{minipage}}\par}} 

\newcommand{\bibit}{\nineit}
\newcommand{\bibbf}{\ninebf}
\renewenvironment{thebibliography}[1]
	{\frenchspacing
	 \ninerm\baselineskip=11pt
	 \begin{list}{\arabic{enumi}.}
	{\usecounter{enumi}\setlength{\parsep}{0pt}
	 \setlength{\leftmargin 12.7pt}{\rightmargin 0pt} 
	 \setlength{\itemsep}{0pt} \settowidth
	{\labelwidth}{#1.}\sloppy}}{\end{list}}

\newcounter{itemlistc}
\newcounter{romanlistc}
\newcounter{alphlistc}
\newcounter{arabiclistc}

\newcommand{\fcaption}[1]{
        \refstepcounter{figure}
        \setbox\@tempboxa = \hbox{\footnotesize Fig.~\thefigure. #1}
        \ifdim \wd\@tempboxa > 5in
           {\begin{center}
        \parbox{5in}{\footnotesize\smalllineskip Fig.~\thefigure. #1}
            \end{center}}
        \else
             {\begin{center}
             {\footnotesize Fig.~\thefigure. #1}
              \end{center}}
        \fi}

\newcommand{\tcaption}[1]{
        \refstepcounter{table}
        \setbox\@tempboxa = \hbox{\footnotesize Table~\thetable. #1}
        \ifdim \wd\@tempboxa > 5in
           {\begin{center}
        \parbox{5in}{\footnotesize\smalllineskip Table~\thetable. #1}
            \end{center}}
        \else
             {\begin{center}
             {\footnotesize Table~\thetable. #1}
              \end{center}}
        \fi}

\def\@citex[#1]#2{\if@filesw\immediate\write\@auxout
	{\string\citation{#2}}\fi
\def\@citea{}\@cite{\@for\@citeb:=#2\do
	{\@citea\def\@citea{,}\@ifundefined
	{b@\@citeb}{{\bf ?}\@warning
	{Citation `\@citeb' on page \thepage \space undefined}}
	{\csname b@\@citeb\endcsname}}}{#1}}

\newif\if@cghi
\def\cite{\@cghitrue\@ifnextchar [{\@tempswatrue
	\@citex}{\@tempswafalse\@citex[]}}
\def\citelow{\@cghifalse\@ifnextchar [{\@tempswatrue
	\@citex}{\@tempswafalse\@citex[]}}
\def\@cite#1#2{{$\null^{#1}$\if@tempswa\typeout
	{IJCGA warning: optional citation argument 
	ignored: `#2'} \fi}}

\def\pmb#1{\setbox0=\hbox{#1}
	\kern-.025em\copy0\kern-\wd0
	\kern.05em\copy0\kern-\wd0
	\kern-.025em\raise.0433em\box0}

\def\fnt#1#2{\footnotetext{\kern-.3em
	{$^{\mbox{\scriptsize #1}}$}{#2}}}

\def\thefootnote{\fnsymbol{footnote}}
\def\@makefnmark{\hbox to 0pt{$^{\@thefnmark}$\hss}}	
	
\def\ps@myheadings{%
    \let\@oddfoot\@empty\let\@evenfoot\@empty
    \def\@evenhead{\slshape\leftmark\hfil}
    \def\@oddhead{\hfil{\slshape\rightmark}}
    \let\@mkboth\@gobbletwo
    \let\sectionmark\@gobble
    \let\subsectionmark\@gobble
    }
\font\tenrm=cmr10
\font\tenit=cmti10 
\font\tenbf=cmbx10
\font\bfit=cmbxti10 at 10pt
\font\ninerm=cmr9
\font\nineit=cmti9
\font\ninebf=cmbx9
\font\eightrm=cmr8

\def\qed{\hbox{${\vcenter{\vbox{			
   \hrule height 0.4pt\hbox{\vrule width 0.4pt height 6pt
   \kern5pt\vrule width 0.4pt}\hrule height 0.4pt}}}$}}

\renewcommand{\thefootnote}{\fnsymbol{footnote}}  

\begin{document}
\setlength{\textheight}{7.7truein}  

\thispagestyle{empty}

\markboth{\protect{\footnotesize\it Correlation between plasma and temperature
corrections to the Casimir force}}{\protect{\footnotesize\it Correlation between 
plasma and temperature corrections to the Casimir force}}

\normalsize\textlineskip

\setcounter{page}{1}


\vspace*{0.88truein}

\centerline{\bf CORRELATION BETWEEN PLASMA AND TEMPERATURE}
\vspace*{0.035truein}
\centerline{\bf CORRECTIONS TO THE CASIMIR FORCE}
\vspace*{0.37truein}
\centerline{\footnotesize Cyriaque GENET, Astrid LAMBRECHT and Serge REYNAUD 
\footnote{mailto:genet@spectro.jussieu.fr, 
lambrecht@spectro.jussieu.fr, reynaud@spectro.jussieu.fr; \\
http://www.spectro.jussieu.fr/Vacuum}}
\baselineskip=12pt
\centerline{\footnotesize\it Laboratoire Kastler Brossel
\footnote{Laboratoire du CNRS, de l'Ecole Normale Sup\'{e}rieure 
et de l'Universit\'{e} Pierre et Marie Curie} , UPMC case 74 }
\baselineskip=10pt
\centerline{\footnotesize\it Campus Jussieu, F-75252 Paris Cedex 05, France}

\vspace*{0.225truein}

\vspace*{0.21truein}
\abstracts{When comparing experimental results with theoretical predictions 
of the Casimir force, the accuracy of the theory is as important as the
precision of experiments. Here we evaluate the Casimir force when finite 
conductivity of the reflectors and finite temperature are simultaneously 
taken into account. We show that these two corrections are correlated, 
{\it i.e.} that they can not, in principle, be evaluated separately and 
simply multiplied. We estimate the correlation factor which
measures the deviation from this common approximation.
We focus our attention on the case of smooth and plane plates with a 
metallic optical response modeled by a plasma model. }{}{}

\setcounter{footnote}{0}
\renewcommand{\thefootnote}{\alph{footnote}}

\vspace*{1pt}\textlineskip	
\section{Motivations}	
\vspace*{-0.5pt}
\noindent
After its prediction in 1948 \cite{Casimir48}, the Casimir force has been observed in a number 
of `historic' experiments \cite{Sparnaay89}. It has recently been measured with an 
improved experimental precision \cite{Bordag01}. The recent experiments should allow for an 
accurate comparison between the measured force and the theoretical prediction 
and this is important for at least two reasons. 

First, accurate experiments are devoted to searches for hypothetical new forces predicted by 
theoretical unification models with nanometric to millimetric ranges 
\cite{Bordag01,Carugno97,Fischbach99,Long99} or by tests of Newtonian gravity at millimetric 
to centimetric distances \cite{Fischbach98}. At submillimetric distances, the Casimir 
effect dominates the hypothetical new force so that the latter would appear as a difference 
between experimental measurements and theoretical expectations of the Casimir force. 

Then, the Casimir force is the most accessible effect of vacuum fluctuations 
in the macroscopic world. As the existence of vacuum energy raises difficulties 
at the interface between the theories of quantum and gravitational phenomena, 
it is worth testing this effect with the greatest care and highest accuracy 
\cite{Reynaud01}. Now, as far as a theory-experiment comparison is concerned, the 
accuracy of theory is as crucial as the precision of experiments. If a given accuracy, 
say at the $1 \%$ level, is aimed at in the comparison, then the theory as well as 
the experiment must be mastered at this level independently from each other.

The differences between the real experimental conditions and the ideal 
situation considered by Casimir play a key role in this discussion. 
Casimir calculated the force between perfectly plane, flat and parallel  
plates in the limit of zero temperature and perfect reflection. 
This is the reason why the Casimir formula $F_{\rm Cas}$ only depends 
on the distance $L$, the area $A$ (supposed to be much larger than $L^2$) and 
the two fundamental constants $c$ and $\hbar$
\begin{eqnarray} 
F_{\rm Cas} &=&\frac{\hbar c A \pi ^2}{240L^4}   \label{Fcasimir} 
\end{eqnarray}
This is a remarkably universal feature, especially since the force is independent 
of the fine structure constant in contrast to the Van der Waals forces. This
indicates that the response of perfect mirrors to the fields is saturated, since 
they reflect 100 \% of the incoming light. 

But experiments are performed with mirrors which do not reflect perfectly the field 
at all frequencies. For example, conduction electrons have an optical response 
described by a plasma model so that metallic mirrors show perfect 
reflection only at frequencies smaller than the plasma frequency $\omega _{\rm P}$. 
Hence the Casimir force between metal plates does fit the Casimir formula 
(\ref{Fcasimir}) only at distances $L$ larger than the plasma wavelength  
$\lambda _{\rm P}=\frac{2\pi c}{\omega _{\rm P}} $.
For metals used in the recent experiments, this wavelength lies in the 
$0.1 \mu$m range ($\sim$ 107 nm for Al and 136 nm for Cu and Au).
At distances smaller than or of the order of $\lambda _{\rm P}$, the finite 
conductivity of the metal produces a reduction of the force  
which can be described by a plasma correction factor \cite{Lambrecht00} 
\begin{eqnarray} 
F^{\rm P} = \eta _{\rm F}^{\rm P} F_{\rm Cas}  \qquad  \eta _{\rm F}^{\rm P} < 1
\end{eqnarray}

At the same time, experiments are performed at room temperature and the radiation
pressure of thermal field fluctuations is superimposed to that of vacuum field
fluctuations. This effect can be described by a temperature correction factor 
which increases the Casimir force at distances larger than or of 
the order of a thermal wavelength $\lambda _{\rm T} = \frac{\hbar c}{k_{\rm B}T}$ 
($\sim 7\mu$m at room temperature)  
\begin{eqnarray} 
F^{\rm T} = \eta _{\rm F}^{\rm T} F_{\rm Cas}  \qquad  \eta _{\rm F}^{\rm T} > 1
\end{eqnarray}

Now, the plasma correction $\eta _{\rm F}^{\rm P}$ has been defined at 
zero temperature while the thermal correction $\eta _{\rm F}^{\rm T}$ 
is usually computed for perfect reflection. Since the plasma wavelength
is much smaller than the thermal wavelength, 
the whole correction $\eta _{\rm F}$ giving the force $F$
when both effects are simultaneously accounted for is commonly calculated 
by multiplying the plasma and thermal correction factors. 
This is however an approximation and the discussion of its accuracy is the main 
motivation of the present paper. To this aim, we write the whole correction factor as
\begin{eqnarray} 
F &=& \eta _{\rm F} F_{\rm Cas} \qquad \qquad 
\eta _{\rm F} = \eta _{\rm F}^{\rm P} \eta _{\rm F}^{\rm T}
\left( 1 + \delta_{\rm F} \right)   \label{deltaF} 
\end{eqnarray}
A null value for $\delta_{\rm F}$ would justify the common approximation where the plasma 
and thermal corrections are computed independently from each other and then multiplied. 
In contrast, a non null value represents a correlation of the two corrections which must be
taken into account in an accurate evaluation of the Casimir force.        

In the present paper, we consider the initial Casimir geometry with perfectly plane, flat 
and parallel plates and thus restrict our attention on conductivity and thermal corrections. 
Since the correlation between these two corrections is appreciable only at distances where 
the plasma model is a good description of metals, we focus our attention on this model
(see a more precise argument below). 

\section{The plasma and thermal corrections to the Casimir force}

\noindent We now present the evaluation of the correction factors 
in the Casimir geometry. 

A cavity built on partly transmitting mirrors can be dealt with using the Fabry-Pérot theory.  
Field fluctuations impinging the cavity have their energy either enhanced or decreased 
inside the cavity, depending on whether their frequency is resonant or not with a cavity mode. 
The radiation pressure associated with these fluctuations then exerts a force on the mirrors
which is directed either outwards or inwards respectively. It is the balance between 
these outward and inward contributions, integrated over the wavevectors associated with 
the field modes, which gives the net Casimir force \cite{Jaekel91}.  

The obtained expression is a generalization of Lifshitz's formula \cite{Lifshitz56} which 
is valid for any couple of mirrors described by arbitrary frequency dependent 
reflection amplitudes obeying the general properties of scattering theory 
\footnote{Lifshits's results were not originally written in terms of reflection coefficients.
U. Mohideen and V.M. Mostepanenko have recently drawn our attention to the reference 
\cite{Kats77} where Kats wrote Lifshits's results in this manner.} .
Since any real mirror is transparent at the high frequency limit, a regular expression 
is naturally obtained, which is free from the divergencies usually associated with 
the infiniteness of vacuum energy.

At non zero temperature, the Casimir force may be written as the Poisson formula given 
by equation (7) in \cite{Genet00}. This formula is used as the starting point of 
the following calculations after specialization to the 
case of metallic mirrors with an optical response described by the plasma model
\footnote{Thermal corrections to the Casimir force have also been evaluated with
a Drude model used to describe absorption in the metal and they have led to 
controversial results \cite{Bostrom00,Svetovoy00,Bordag00}. As far as this controversy is
concerned, note that equation (17) of \cite{Bordag00} coincides with the Poisson formula 
used in the present paper (equation (7) of \cite{Genet00}) and leads to results at 
variance with the ones obtained in \cite{Bostrom00,Svetovoy00}.
For a detailed discussion of the interplay between metallic and temperature 
corrections in the general case, see the contribution of G.L. Klimchitskaya 
to this volume \cite{Klimchitskaya01}.} . 

\begin{figure}[htbp] 
\vspace*{13pt} 
\centerline{\psfig{file=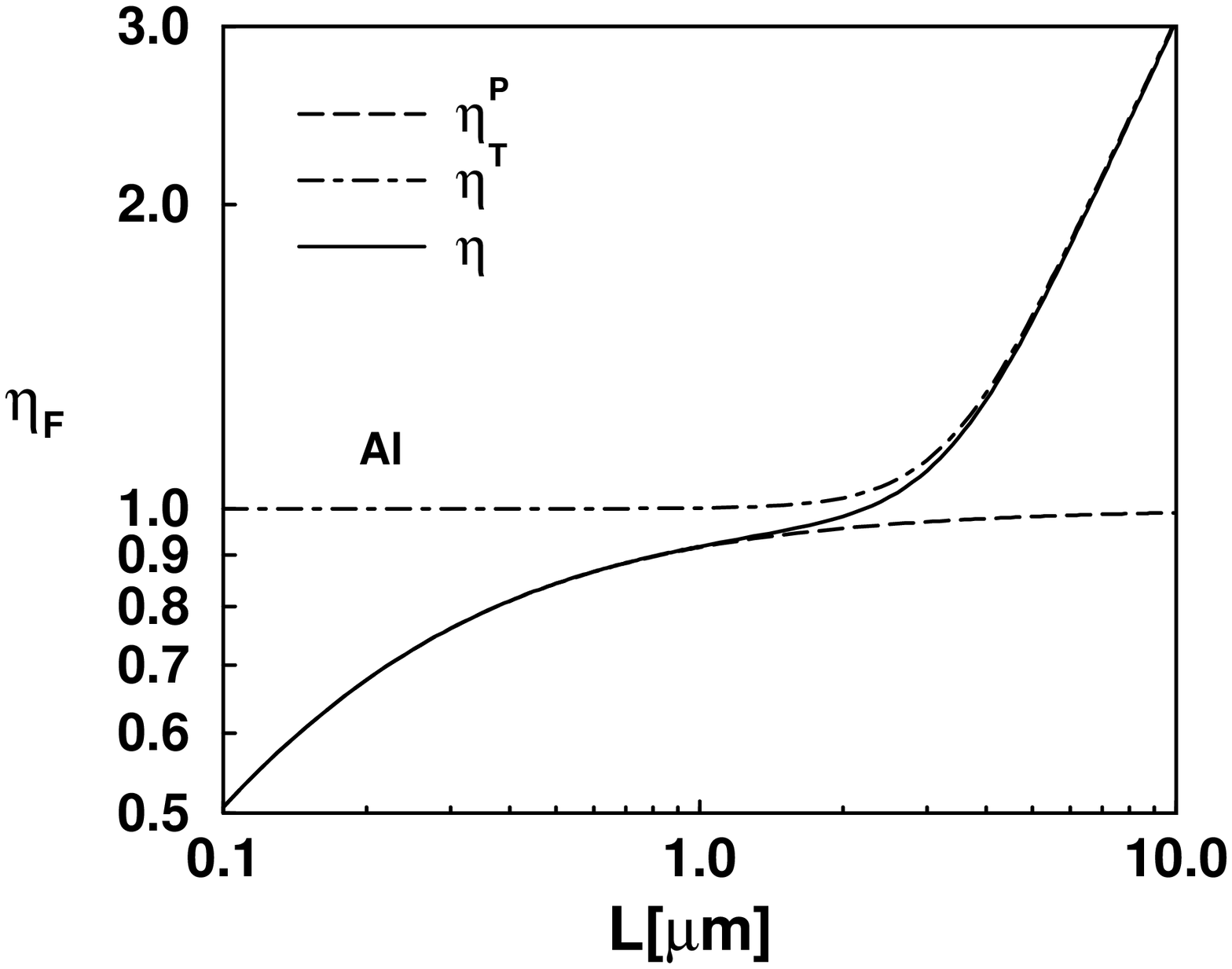,width=7cm}} 
\vspace*{13pt} 
\fcaption{Correction factors for the Casimir force between Al mirrors (plasma model with
$\lambda_{\rm P}=107$nm) at room temperature ($T=300$K) as functions of the distance $L$.
The solid, dashed and dotted-dashed lines correspond respectively to the global correction 
factor $\eta _{\rm F}$, the plasma correction factor $\eta _{\rm F}^{\rm P}$ and the 
thermal correction factor $\eta _{\rm F}^{\rm T}$.} 
\label{fig1} 
\end{figure}

We have numerically evaluated the Casimir force with the plasma wavelength corresponding to 
Aluminium and at room temperature. The global correction factor obtained in this manner is 
shown on figure \ref{fig1} for the experimentally relevant distance range $0.1-10 \mu$m and 
it is compared with the plasma correction factor (evaluated at zero temperature) and the 
thermal correction factor (evaluated for perfect mirrors). 

The thermal correction is negligible at short distances and enhances the force
at large distances whereas the conductivity correction may be ignored at large distances and
decreases the force at small distances. As a consequence, the global correction factor 
$\eta_{\rm F}$ behaves roughly as the product $\eta _{\rm F}^{\rm P} \eta_{\rm F} ^{\rm T}$ 
of the two correction factors evaluated separately. 
But both corrections are appreciable in the intermediate distance range and it is therefore
necessary to discuss more precisely the accuracy of the decorrelation approximation.  

\section{The correlation between correction factors}

\noindent In order to assess the quality of the decorrelation approximation, 
we now evaluate the correlation factor $\delta _{\rm F}$ introduced in (\ref{deltaF}). 

\begin{figure}[htbp] 
\vspace*{13pt}  
\centerline{\psfig{file=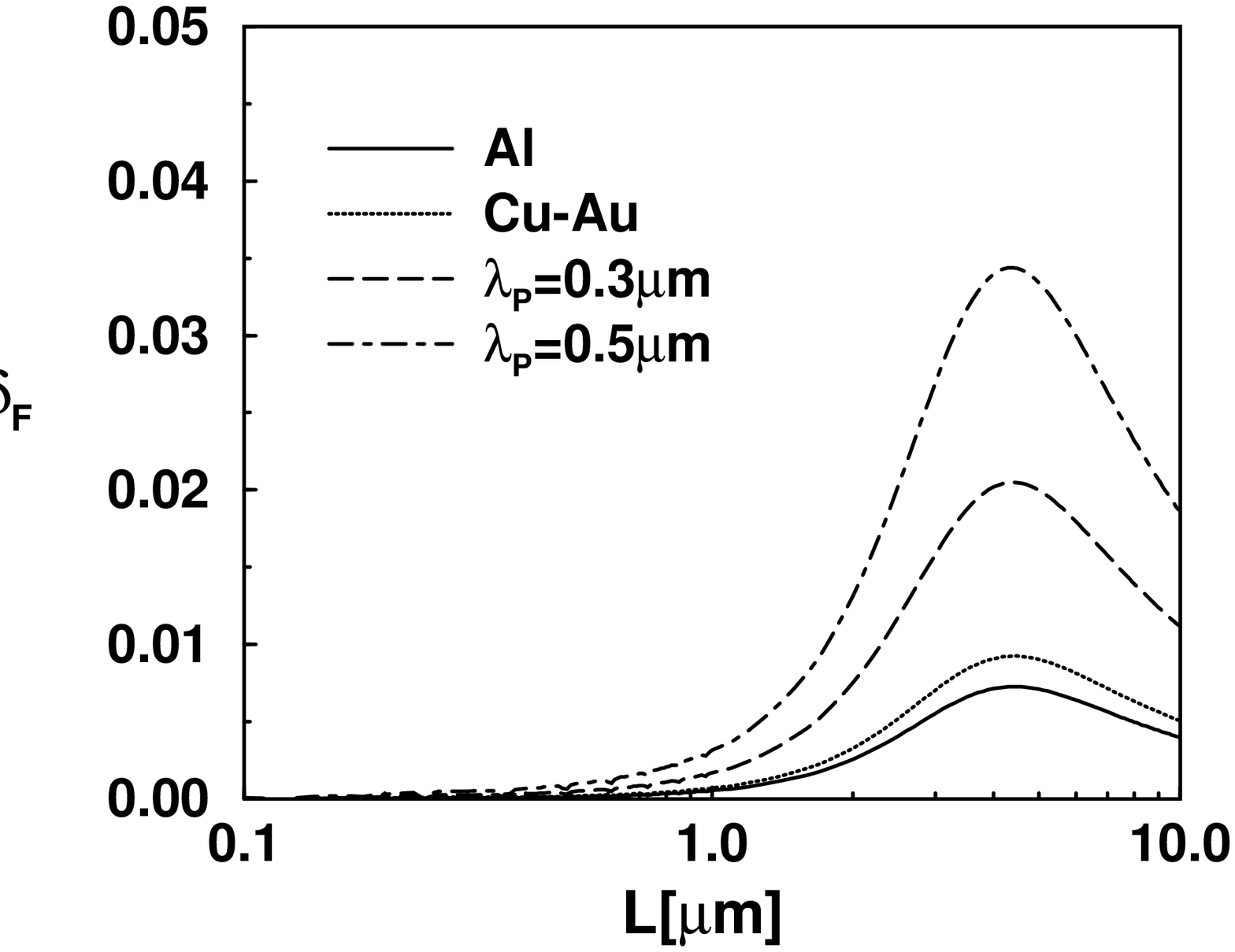,width=7cm}}   
\vspace*{13pt}  
\fcaption{Correlation factor $\delta _{\rm F}$ between the plasma and thermal corrections
as a function of the mirrors distance $L$. The results are given for the plasma wavelengths 
corresponding to Al (107nm), Cu-Au (136nm) and two larger values (300 and 500nm).}  
\label{fig2}  
\end{figure}

This factor is plotted on figure \ref{fig2} as a function of the distance $L$
for the plasma wavelengths corresponding respectively to Al, Cu-Au, and two additional 
plasma wavelengths  chosen to emphasize the correlation effect. 
It turns out that the correlation factor lies in the $\%$ range for Al, Cu-Au, that is 
precisely the accuracy which is usually aimed at. 
The positive sign obtained for $\delta_{\rm F}$ means that the correlation 
increases the theoretical value of the force.  

This is the main result of the present contribution: approximating the global correction factor
as the product of the plasma and thermal correction factors evaluated independently is sufficient
for rough estimates, with a precision worse than $1\%$. But the correlation $\delta_{\rm F}$ 
between these two corrections should be taken into account when an accuracy beyond the $1\%$ level is
needed. At this point, let us emphasize that the correlation is appreciable at distances 
larger than 1$\mu$m where the plasma model is known to be a good effective description of 
the metallic optical response \cite{Lambrecht00}. This justifies the use of this model 
in the present paper which is devoted to the study of the correlation effect
\footnote{At shorter distances, say around 0.1-0.5$\mu$m, a more complete description of the metallic 
optical response must be used in order to obtain accurate estimates of the Casimir force 
\cite{Lambrecht00}. Since the temperature correction is negligible in this distance range,
the estimation can be simplified by considering only the contribution of vacuum fluctuations.} .

Figure \ref{fig2} also shows that the correlation factor is proportional to the value of the plasma
wavelength while keeping the same functional dependence on the distance. It has been
possible to prove that $\delta_{\rm F}$ obeys a simple scaling law \cite{Genet00} 
\begin{equation} 
\delta_{\rm F} = \frac{\lambda _{\rm P}}{\lambda _{\rm T}} \Delta_{\rm F}  \label{Scaling} 
\end{equation} 
It is proportional on one hand to the ratio $\frac{\lambda _{\rm P}}{\lambda _{\rm T}} $ 
of the two wavelengths which characterize respectively the plasma and thermal effects and, 
on the other hand, to the universal function $\Delta _{\rm F}$ which does only depend 
on $\frac {L}{\lambda _{\rm T}}$. This scaling law is valid for 
$\lambda _{\rm P} \ll \lambda _{\rm T}$, which is the situation of interest
for experiments with ordinary metals at room temperature.

An analytical derivation of (\ref{Scaling}) has been given in \cite{Genet00} through a perturbative 
development of the force to first order in $\frac{\lambda _{\rm P}}{\lambda _{\rm T}} $. 
The resulting expression is found to fit well the results of the complete numerical integration  
presented above, with an accuracy now much better that the $1\%$ level. It provides one with 
a simple method for getting an accurate theoretical expectation of the Casimir force in presence 
of plasma and thermal corrections: the force is indeed given by equations (\ref{deltaF},
\ref{Scaling}) and the analytical expressions of $\eta _{\rm F}^{\rm P}$, 
$\eta_{\rm F} ^{\rm T}$ and $\Delta _{\rm F}$ available from \cite{Genet00}.  
This solves the problem of the accurate evaluation of the Casimir force between 
two metallic planes at room temperature at distances where the plasma model can be used.

When addressing the problem of accuracy of theoretical predictions, we must keep in mind 
other corrections involved in recent measurements of the Casimir force \cite{Bordag01}, 
in particular the geometry correction - experiments are not performed with two plane plates 
but with a sphere and a plane - and the roughness correction.
This entails that not only the accuracy of the approximations used to treat these effects 
should be carefully studied for perfect mirrors in vacuum but also that the correlations 
between geometry, roughness, conductivity and temperature corrections have to be evaluated.

\nonumsection{Acknowledgements}
\noindent
Thanks are due to G. Barton, M. Bordag, R. Esquivel-Sirvent, C. Farina, G.L. Klimchitskaya, 
P.A. Maia Neto, U. Mohideen, V.M. Mostepanenko and C. Villarreal for stimulating discussions 
which have inspired several additions to the present contribution.


\end{document}